\documentclass[12pt]{elsarticle}

\usepackage{graphics}
\usepackage{graphicx}
\usepackage{epsfig}
\usepackage{amssymb}
 \usepackage{amsthm}

\newcommand{\kmps}{kms$^{-1}$}
\newcommand{\vicme}{V$_{\rm ICME}$}
\newcommand{\vint}{V$_{\rm INT}$}
\newcommand{\vips}{V$_{\rm SHOCK}$}
\newcommand{\vest}{V$_{\rm EST}$}
\newcommand{\vavg}{V$_{\rm AVG}$}
\newcommand{\mpsq}{ms$^{-2}$}

\journal{JASTP}

\begin{document}

\begin{frontmatter}

\title{Coronal Mass Ejections - Propagation Time and Associated Internal Energy}

\author[]{P.K. Manoharan$^a$ and A. Mujiber Rahman$^{b,}$\fnref{1}}
\address[$^1$]{Radio Astronomy Centre, National Centre for Radio Astrophysics,\\
               Tata Institute of Fundamental Research, Udhagamandalam (Ooty) 
               643001, India.}

\address[$^2$]{School of Physics, Madurai Kamaraj University, Madurai 625021, India.}
\fntext[a]{on leave from Department of Physics, Hajee Karutha Rowther Howdia College,
Uthamapalayam 625533, India.}

\begin{abstract}

In this paper, we analyze 91 coronal mass ejection (CME) events studied by 
Manoharan et al.~(2004) and Gopalswamy and Xie (2008). These earth-directed
CMEs are large (width $>$ 160$^\circ$) and cover a wide range of speeds 
($\sim$120--2400 {\kmps}) in the LASCO field of view. This set of events
also includes interacting CMEs and some of them take longer time to reach 
1 AU than the travel time inferred from their speeds at 1 AU. We study the link 
between the travel time of the CME to 1 AU (combined with its final speed
at the Earth) and the effective acceleration in the Sun-Earth distance. 
Results indicate that (1) for almost all the events (85 out of 91 events), 
the speed of the CME 
at 1 AU is always less than or equal to its initial speed measured at the 
near-Sun region, (2) the distributions of initial speeds, CME-driven shock 
and CME speeds at 1 AU clearly show the effects of aero-dynamical drag 
between the CME and the solar wind and in consequence, the speed of the CME 
tends to equalize to that of the background solar wind, (3) for a large
fraction of CMEs (for $\sim$50\% of the events), the inferred effective 
acceleration along the Sun-Earth line dominates the above drag force. The 
net acceleration suggests an average dissipation of energy $\sim$10$^{31-32}$ 
ergs, which is likely provided by the Lorentz force associated 
with the internal magnetic energy carried by the CME. 

\end{abstract}


\end{frontmatter}


\section{Introduction}

It is now well established that the major non-recurrent geomagnetic storms 
are generally associated with the solar wind disturbances generated by fast 
coronal mass ejections (CMEs). The intensity of such storms primarily depends 
on the CME size, speed, density, and strength of the southward magnetic field 
component at the near-Earth space (e.g., Tsurutani et al., 2006).
Another aspect of CME-related space weather 
is large CME-driven interplanetary disturbances that are usually preceded by 
strong shocks and they are effective accelerators of particles as well as 
sources of radio emissions. There is a great interest in understanding the 
propagation characteristics of CMEs and predicting the CME-related space 
weather. In this connection several studies have been made to investigate 
(a) the relationship between the speed of the CME at the near-Sun region and 
effects created by its shock ejecta and/or magnetic cloud at the Earth's 
magnetosphere (e.g., Richardson et al., 2007), 
and (b) the arrival times of the CME and its associated shock at 1 AU 
(e.g., Gopalswamy et al., 2001; 2005; Manoharan et al., 2004). 
In such recent studies, white-light images mainly obtained from 
LASCO/SOHO space mission and 1-AU in-situ measurements have been employed.

For example, Gopalswamy et al.,~(2000) found that in the Sun-Earth distance, 
CMEs experienced an effective acceleration, which was highly correlated with 
the initial speed of the CME. They proposed a kinematical model based on the 
single parameter, i.e., the initial speed of the CME, to predict the arrival 
time of CME. In subsequent studies (e.g., Gopalswamy et al., 2001, 2005; 
Manoharan et al., 2004; Michalek et al., 2004), the above model got refined with an 
acceleration-cessation distance. Since the exact distance at which the 
effective acceleration ceased was not known, a particular acceleration-cessation 
distance of 0.76 AU was assumed. However, these empirical models suffer the problem 
of projected speed of the CME in the coronagraph field of view. Since the initial 
speed of the CME obtained from the white-light images is in the sky-plane, the 
speed along the Sun-Earth line is required to predict the travel time of the CME. 
Therefore, a cone-shaped structure of the CME, instead of the simple geometrical 
correction,
can be assumed to obtain the radial speed of the CME, which then used as an input to 
the empirical shock arrival (ESA) model to obtain the Sun to Earth travel time of 
shocks (Michalek et al., 2004; Xie et al., 2006). The earthward speed employed in these
empirical models predicted the travel time within about 75\% accuracy (Gopalswamy 
et al.,2008; Cho et al., 2003, Xie et al., 2006).

Recent studies have however further shown that the radial profile of speed can 
differ from one CME to the other, depending on the physical properties of the 
CME (i.e., speed, mass, size, and internal magnetic energy) and the background 
solar wind in which CME is propagating (e.g., Manoharan 2006, 2010). Additionally,
the interaction of a CME with preceding CME(s) can also slow down its propagation 
and it tends to take longer time to travel to the Earth (i.e., Manoharan et al., 
2004). Moreover studies on the evolution of CMEs between Sun and 1 AU, using 
white-light images and interplanetary scintillation data demonstrated that the 
internal energy of the CME dominates and contributes to the propagation out to 
$\sim$0.35-0.45 AU (e.g., Manoharan et al., 2001, 2002; 
Joshi et al., 2007; Kumar et al., 2010; Manoharan, 2010). The CME evolution model 
also suggests that the driving Lorentz force caused by the toroidal current 
overcomes the aero-dynamical drag force applied by the background solar wind 
within about ~0.4 AU (e.g., Chen, 1996). It is likely that the effect of Lorentz force, 
which depends on the magnetic flux system embedded within the CME, has direct link 
to the source region of CME on the Sun.

The consequences of the shock associated with a CME at the Earth (e.g., interplanetary 
shock strength, sudden commencement, and/or sudden impulse) are directly linked to 
the energy available within the CME. The combined effects of (a) shock strength, 
(b) magnetic energy within the magnetic cloud, (c) kinetic energy of the 
ejecta, and (d) physical properties of the background solar wind
determine the onset and the intensity of the storm at the 
Earth's magnetosphere. Therefore the essential part is to understand the 
relationship between the shock and its driver CME in terms of CME initial and
final speeds and travel time over a given distance along the Sun-Earth line.
Such an investigation can provide the 
insight on energy lost/gained by the CME on its way from Sun to 1 AU. For example, 
a kinematically and/or magnetically weak CME is likely to lose energy on its way and 
it may not possess sufficient energy to drive a shock ahead of it. In such a situation 
the shock is likely to detach from the driver CME and dissipate with further 
time/distance of propagation.

In this paper, we analyze CMEs studied by Manoharan et al., (2004) and Gopalswamy 
and Xie, (2008) to investigate the link between the travel time of the CME to 1 AU
and its initial speed, final speed (also its associated shock speed) at 1 AU, and
its effective acceleration in the Sun-Earth distance. We also consider the 
distributions of CME initial speed, CME and shock speeds at 1 AU. 
The paper is organised as follows. Section 2 describes the Earth-directed CME
events selected for the study and discusses the correlation between the initial speed 
of the CME in the sky plane and projection corrected speed. In the next section, the 
speed of the CME in the near-Earth space has been compared with its travel time in 
the inner heliosphere. Section 4 provides an account on the average speed of the CME
along the Sun-Earth line and discusses the possible internal energy of the CME
event. The effective acceleration experienced by the CME events is discussed in
Section 5 and the final section gives the summary and conclusion. 

\begin{figure}
\centerline{\includegraphics[width=0.22\textheight,clip=,angle=-90]{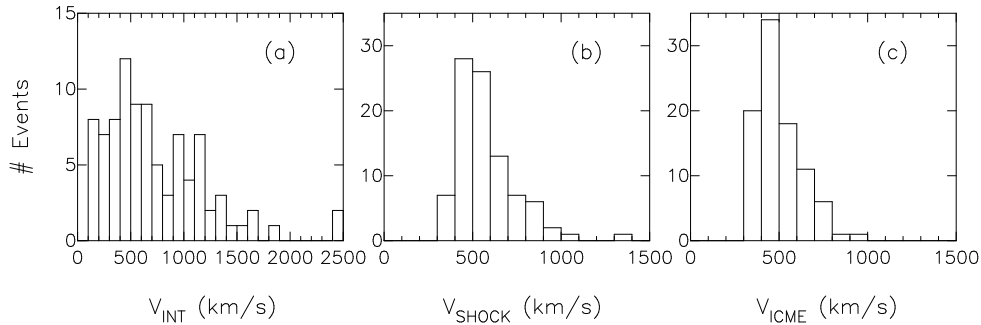}}
\caption{Histograms of (a) initial speed of the CME ({\vint}) in the LASCO field of 
view, (b) speed of the associated interplanetary shock at 1 AU ({\vips}), and
(c) speed of the CME at 1 AU ({\vicme}). It is clearly seen that the wide range of 
speeds in the LASCO field of view, i.e., $\sim$100 -- 2400 {\kmps}, tends to narrow to
a range of $\sim$400--1000 {\kmps} at the near-Earth region in the cases of shock and CME.} 
\end{figure}

\section{List of Earth-directed CME Events}

Manoharan et al., (2004) studied the propagation characteristics of 91 earth-directed 
CME events, originated close to the central meridian of the Sun (within about 
$\pm$30$^\circ$
of heliographic latitude and longitude). These CMEs were wide (i.e., halo and 
partial halo events of width $>$160$^\circ$) and accompanied by 
clear and relatively intense shocks. For each event, the onset of the CME on the Sun 
was identified by analysing EIT and LASCO images and the arrival of CME-shock pair 
at 1 AU was examined using the 
solar wind density, speed, temperature, and interplanetary magnetic field features 
of the ejecta and magnetic cloud (e.g., low-proton temperature, high-charge state of 
iron, strong magnetic field, and systematic rotation in field). In the LASCO field of 
view, these CMEs covered a speed range of {\vint}~$\sim$~120--2400 {\kmps}, with an average of 
$\sim$730 {\kmps}. At 1 AU, Alfvenic Mach numbers of their associated shocks covered a 
range 
between 1.1 and 9. The above list included 25 interacting CME events. The criteria to 
identify the interacting cases were (i) the main fast CME and the preceding slow CME 
originated from the same active region on the Sun and (ii) the slow CME took off within 
a day before the main event. The average speed of $\sim$975 {\kmps} of these interacting 
cases suggests that the subset of fast CMEs could catch up with the slow ones moving 
ahead.

Figure 1 shows the histograms of (a) initial speeds of CMEs in the LASCO field of 
view ({\vint}), (b) in-situ speeds of their corresponding interplanetary shocks ({\vips}) and 
(c) interplanetary CMEs at 1 AU ({\vicme}). At the orbit of the Earth, the speed 
range of these CME events tends to narrow towards the ambient solar wind (i.e.,speed 
$\sim$450--500 {\kmps}). It suggests (1) the effectively slowing down of events resulting
because of the interaction between the background solar wind and the CME and (2) the
exchange of energy between the CME and ambient solar wind.

\begin{figure}[t]
\centerline{\includegraphics[width=0.5\textwidth,clip=,angle=-90]{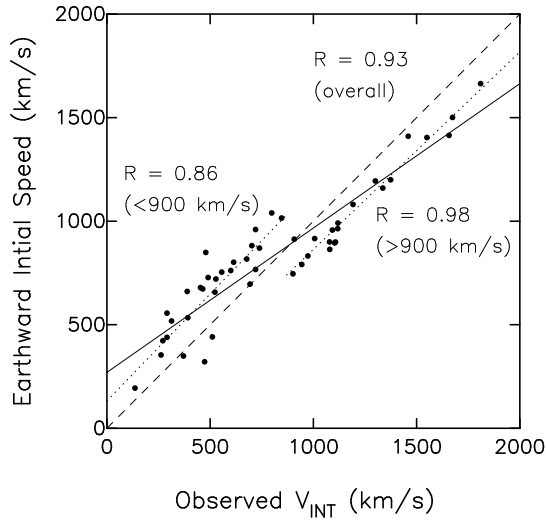}}
\caption{The initial speed of the CME ({\vint}) in the sky-plane as observed 
in the LASCO field of view is plotted against its corrected earthward speed as
obtained by Gopalswamy and Xie (2008). The dashed line represents the 100\% 
correlation line. The continuous line is the overall best fit to all the data 
points and provides a correlation coefficient of R$\sim$93\%. The dotted lines 
are, respectively, best fits to groups of {\vint}$<$900 and {\vint}$>$900 {\kmps} 
data points and their correlation coefficients are also shown. They show
systematic offsets with respect to the 100\% correlation line (refer to text). }
\end{figure}

\subsection{Sky-plane and Earthward Speeds}

Since the LASCO images provide the sky-plane speed of the CME, Gopalswamy and
Xie (2008) computed the speed along the Sun-Earth line. It was found that when 
the projection was taken into consideration, their empirical shock arrival (ESA)
model predicted the arrival of the CME within about 12 hours of uncertainty (also refer 
to Gopalswamy et al., 2000; 2005). The earthward speeds available for 49 out of
91 events from the list given by Gopalswamy and Xie (2008) have been compared 
with their simple initial sky-plane speeds obtained from the LASCO images. 
As shown in Figure 2, the overall correlation between the initial sky-plane
speed and the projection-corrected speed is good (i.e., correlation coefficient 
of R~$\approx$93\%) and the best fit is shown by a continuous line. However,
we see a marked difference between two groups of CMEs having speeds in the range,
(i) {\vint}$<$900 {\kmps} and (ii) {\vint}$>$900 {\kmps}. In the high-speed side
the scatter is less between the sky-plane speed and projection-corrected speed
and the correlation is high, $\sim$98\%. Moreover, the earthward speed shows a
systematic downward offset of $\sim$90 {\kmps} (shown by a dotted line at
{\vint}$>$900 {\kmps}) with respect to the sky-plane speed. Whereas in the 
low-speed side ($<$900 {\kmps}), we see an offset in the positive side, $\sim$130 
{\kmps} ({\vint}$<$900 {\kmps}) and on the average earthward speed is higher
than the sky-plane observed speed. In the low-speed side, we see more scatter
and the correlation is $\sim$86\%.

Apart from the above discussed 49 disk-centered events, the list given by 
Gopalswamy and Xie (2008) also included 23 events, which mostly originated 
at heliographic latitude and longitude $>$$|$$\pm$30$^\circ$$|$. The comparison of 
sky-plane and earthward speeds of these non-disk-centered events also shows 
the downward offset of $\sim$100--150 {\kmps} with respect to the sky-plane 
speeds over the entire speed range. The systematic opposite offsets seen,
respectively, in the low and high speed regions of the disk-centered events 
require more investigations. However, for these events the overall agreement 
between the sky-plane and projection-corrected speeds is good (i.e., R$\sim$93\%, 
refer to Figure 2) and in this study, we consider only the initial sky-plane 
speed obtained from the LASCO images.  

\begin{figure}[t]
\centerline{\includegraphics[width=0.5\textwidth,clip=,angle=-90]{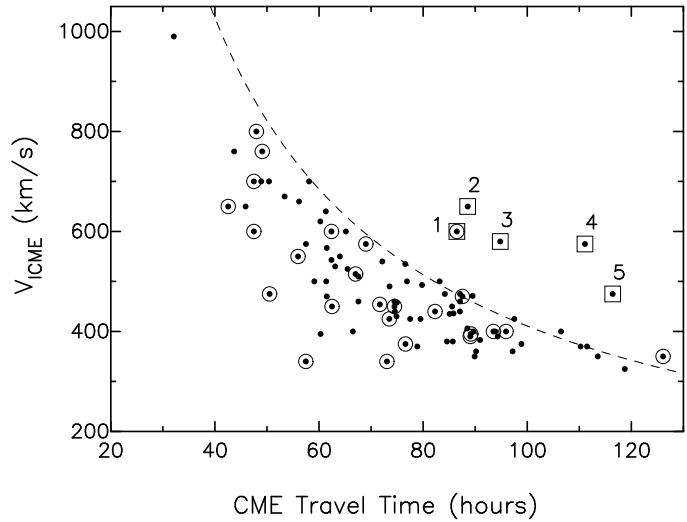}}
\caption{The observed in-situ speed of the CME at the Earth's orbit is plotted 
as a function of travel time of the CME between the LASCO field of view and 1 AU. 
The dashed-line 
curve shows the estimated travel time assuming that the CME propagated with
a constant speed equal to that of its speed at 1 AU ({\vicme}). The points
over-plotted with circle symbol are interacting CMEs. Most of the events 
(except five events shown by square symbols) lie below the assumed-constant 
speed curve.}
\end{figure}

\section{CME Speed at 1 AU}

In Figure 3, the speed of the CME at 1 AU ({\vicme}) has been plotted against 
the travel time 
of the CME between Sun and Earth. The interacting events have been over-plotted with
circle symbols. The continuous dashed line shown in the figure is estimated travel 
time based on the assumption that the propagation speed of the CME event remained 
constant at {\vicme} in the Sun-Earth distance. It is evident that most of the 
events fall below the constant-speed curve. It suggests that on the average, these
CMEs have travelled faster than the final speed at 1 AU ({\vicme}). Those events 
lying on the continuous curve or close to it, indicate a nearly-constant speed of 
propagation in the Sun-Earth line.  Another important point to be noted in the 
above figure is that the value of ${\rm dV/dt}$ within $\sim$70 hours of travel 
time is larger than that of events falling in the 
range $\sim$70-130 hours. The high-speed events ({\vicme} $>$ 500 {\kmps})
show a sharp decline in travel time with the final speed. Whereas events propagating close
to the ambient solar wind speed ({\vicme} $\leq$500 {\kmps}) show a rather weak
dependence of the travel time with speed. It is seen that out of 91 events, 
86 events fall close to or below the constant-speed curve. The comparison of
travel time with the constant-speed curve suggests that (1) the initial speed
of the CME event should have been higher than its final speed at 1 AU, (2) the
aero-dynamical drag force acting on the CME tends to slow down the event, and 
(3) the exchange of energy between the CME and the ambient solar wind 
effectively makes the CME to attain the speed of the background solar wind
(refer to Figure 1).

In Figure 3, five events show significant positive time offset from the 
constant-speed curve. These CMEs have particularly taken longer travel times 
than that corresponding to their speeds at 1 AU (i.e., events falling well above
the constant-speed curve are shown by square symbols and marked by numerals 1 to 
5 in Figure 3). In these cases, it is likely that they have traveled a considerable 
portion of the Sun-Earth distance at low speed and a gradual acceleration 
associated with them have possibly made them to attain a speed faster than 
the initial speed. The interaction suffered by the event \#1 (on June 29, 
1999, at 07:31 UT) with the preceding 
CME has likely modified its speed profile and led to a longer travel time 
(e.g., Manoharan et al., 2004). In the case of event \#2 on July 11, 2000, at 13:27 UT,
it has gone through heavy deceleration (-43 {\mpsq}) in the LASCO field of
view. Moreover, the active region associated with this event produced several
large and fast CMEs (including the Bastille Day event of 2000) and it is 
likely that the interaction between CMEs has modified the speed profile in
the inner heliosphere.  The event \#3 corresponds to a CME observed on June 
20, 2000 at 09:10 UT. However, as per the recent revision made in the CME 
catalogue (http://cdaw.gsfc.nasa.gov/CME\_list), it has been found that two 
events of different speeds (455 and 635 {\kmps}) have originated in the same 
location at the same time. Thus, the interaction between these events has
affected the propagation. The longer travel times taken by events \#4 and \#5, 
respectively, correspond to their low initial speeds of 266 and 192 {\kmps} 
in the LASCO field of view.


\section{Travel Time and Average Initial Speed}

The travel time of a CME event to the Earth is an indicator of its typical 
average speed ({\vavg}) between the LASCO field of view and the Earth. 
Therefore using a given observed final speed at 1 AU ({\vicme}), combined 
with its average speed of propagation in the Sun-Earth distance obtained 
from the total travel time, it is possible to estimate the expected initial 
speed of the CME ({\vest}). The average speed of the CME can be given by,
\begin{eqnarray}
{\rm V_{AVG} }  & = & \left( \frac{\sim 1\,\, {\rm AU}}{{\rm CME\,\,\, Travel\,\,\, Time}}\right) 
                    \\ \nonumber \\ \nonumber
       &  = &  \left( \frac{\rm V_{EST} + V_{ICME}}{2} \right). 
\end{eqnarray}
In the above equations, the value of {\vest} can be determined from the
observed quantities {\vicme} and travel time of the CME. Figure 4 displays 
the correlation plot between the estimated initial speed of the CME 
({\vest}) and observed initial (\vint) speed in the near-Sun region 
(i.e., in the LASCO field of view). 

\begin{figure}[t]
\centerline{\includegraphics[width=0.5\textwidth,clip=,angle=-90]{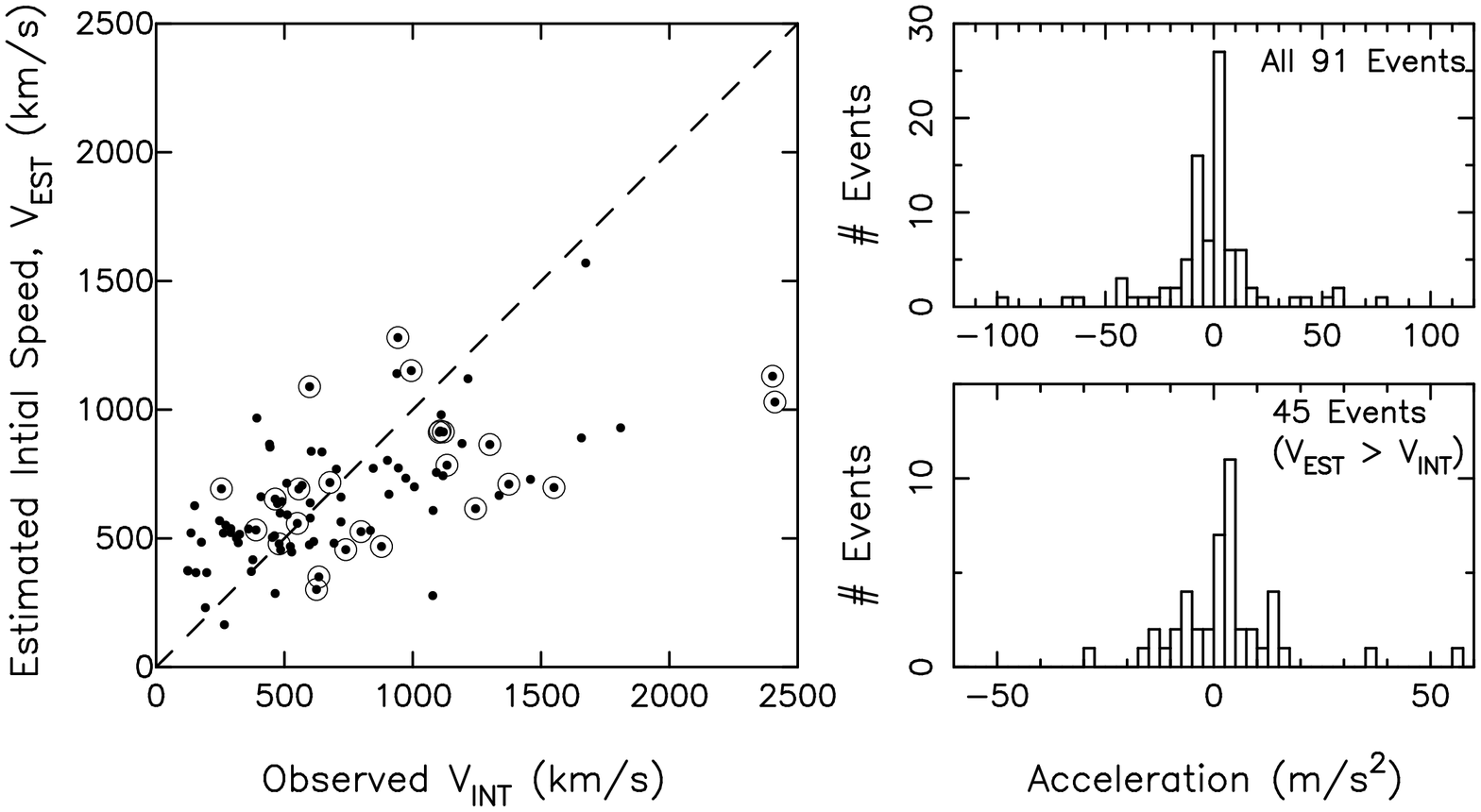}}
\caption{The estimated initial speed ({\vest}, refer to equation 1) obtained 
from the travel time and final speed of the CME at 1 AU ({\vicme}) is compared 
with the observed
initial speed of the CME ({\vint}) in the LASCO field of view. The dashed
line indicates the 100\% correlaton between {\vest} and {\vint}. The events
lying above the line reveal that since {\vest} $>$ {\vint}, these events 
have been supplied with some excess energy for the propagation and they
could overcome the drag force applied by the ambient solar wind. On the
other hand, events lying below the line have suffered deceleration
and taken more travel time as indicated by {\vest} $<$ {\vicme}. The
interacting CME events are marked with circle symbols. On the right-hand side 
of the plot histograms are shown for (a) effective acceleration of all 91 events in
the LASCO field of view and (b) only for 45 events of {\vest} $>$ {\vint}.}
\end{figure}

\subsection{Internal Energy}

It is evident from Figure 4 that on the whole the observed initial speeds 
(\vint) are not well correlated with the estimated speeds ({\vest}). However,
it is interesting to note that nearly 50\% of the events (45 events
out of 91 events) are located above the correlation line, indicating higher
estimated speed than that of the observed initial speed (\vest~$>$~\vint).
These events show an average of $<$\vest$>$ $\sim$ 638 {\kmps}. However, 
it is to be noted that the average of these events' observed initial speeds 
is $<$\vint$>$ $\sim$ 427 {\kmps}, which is considerably lower than 
that of $<$\vest$>$. Since the ambient solar wind speed is considerably lower 
than $<${\vest}$>$, the above result based on the observed travel time (also 
{\vicme})
suggests that these events have been supported by an excess energy associated 
with the CME, i.e., internal magnetic energy, which is revealed in the form 
of higher estimated initial speed ({\vest}). 

If an average mass for the halo CME events is considered, the average initial 
estimated speed ($<${\vest}$>$) provides a typical energy of $\sim$10$^{32}$ 
ergs. It is about 50\% larger than the energy associated with the average 
initial observed speed ($<${\vint}$>$) of the CMEs. Thus, the above exercise 
suggests that the internal magnetic energy in the range of $\sim$10$^{31-32}$ 
ergs stored within the CME has been utilized to overcome the drag force and/or 
transferred to the ambient solar wind. Since the energy to drive a CME phenomenon 
comes from the `free energy' available in the magnetic fields of the active
region, the currents within the field must support the propagation (e.g.,
Metcalf et al., 2008). The typical energy dissipated in the Sun-Earth distance
is consistent with the two-step deceleration of CME speed profile (i.e., a low
or moderate deceleration within $\sim$0.5 AU and a rapid deceleration at further
large distance) and the support of internal magnetic energy (i.e., Lorentz force)
in the propagation of the CME (e.g., Manoharan
et al.,~2001; Manoharan, 2006; Tokumaru et al.,~2000; Howard et al.,~2007; 
Gonzalez-Esparza and Aguilar-Rodriguez, 2009; Chen, 1996).

Whereas the remaining events of lower estimated speed (i.e., {\vest}~$<$~{\vint};
refer to events falling below the correlation line in Figure 4) 
than the observed initial speed show an opposite trend in averages, respectively, 
$<$\vest$>$ $\sim$ 682 {\kmps} and $<$\vint$>$ $\sim$ 1029 {\kmps}. These events 
in fact have started with a higher initial speed (\vint) than the estimated speed 
(\vest). But, the observed initial speed and travel time indicate that these events 
have suffered an effective
longer propagation time. In other words, they could not overcome the drag imposed
on them by the ambient solar wind and likely decelerated in the Sun-Earth distance.
The net effect of simple dragging of the CME in the solar wind (i.e., significantly 
less energy available to overcome the drag force) is indicated by an average lower
estimated initial speed than the observed average initial speed ($<${\vest}$>$ $<$
$<${\vint}$>$).

Additionally, as shown in Figure 4 (right-side panels), the distributions 
of effective accelerations for all 91 events and events having {\vest} $>$ 
{\vint} in the LASCO field of view look nearly similar. Whereas, events 
lying below the correlation line (i.e., remaining 46 events of {\vest} $<$ 
{\vint}) show respectively, the average observed initial speed and the 
average estimated speed of 1029 and 682 {\kmps}. It is inferred that these 
CMEs decelerated heavily due to the interaction between the CME and 
background solar wind. 

Therefore in the group of ``{\vest} $>$ {\vint}'' 
events, the resulting high-estimated speed based on the average speed 
and observed travel time suggests that these events were supplied with 
the excess of energy while propagating from Sun to Earth. In other words, 
their propagation has been supported/aided by the internal energy 
stored within the CME. Numerical simulations and CME propagation studies 
have shown that the Lorentz force play a crucial role in supporting 
the propagation.

\begin{figure}
\centerline{\includegraphics[width=0.4\textwidth,clip=,angle=-90]{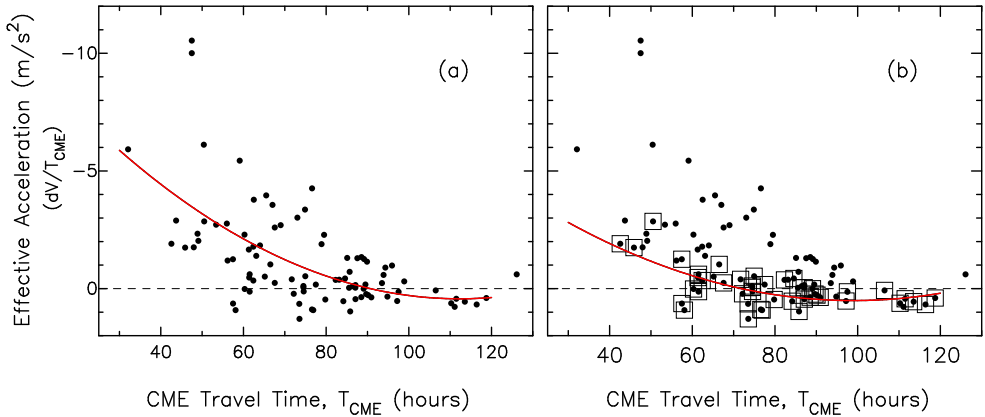}}
\caption{(a) The effective acceleration of the CME as a function of its final
speed at 1 AU. The horizontal line defines the constant speed of propagation
(i.e., net `zero' acceleration). Events lying below the no effective
acceleration line have gone through net positive acceleration and above it
have experienced deceleration. (b) same as the above plot and events of 
{\vest}~$>$~{\vint} are shown by over-plotted square symbols. }
\end{figure}

\section{Effective Acceleration}

The net effective acceleration of an event measured in the Sun-Earth distance 
can be given by the change in the speed of the CME divided by its travel time. 
Figure 5 shows 
the effective acceleration plotted against the travel time of each event. The 
dashed horizontal line shows the zero acceleration line, which indicates that the 
CME traveled at a constant speed. The points lying below this line have gone 
through an effective positive acceleration and events above it have experienced 
a net deceleration. When we compare this plot with Figure 3, we see that the
high-speed CMEs (or events of
small transit time, $<$50 hours) tend to decelerate heavily than that of the 
low-speed events. In Figure 5a, the continuous curve is the least-square fit 
to all the CME events. It provides an average relationship between the effective
acceleration experienced by the CME and its travel time in the Sun-Earth
distance, as given by,
\begin{eqnarray}
 {\rm A_{CME} } = -11.3 + 2.1\times 10^{-1}\, {\rm T_{CME} } - 9.3\times 10^{-4} 
\, {\rm T^{2}_{CME} },
\end{eqnarray}
where A$_{\rm CME}$ is the effective acceleration (given in {\mpsq}) of the
CME and T$_{\rm CME}$ (given in hour) is the travel time of the CME in the 
Sun-Earth distance. 

In principle, a CME event traveling with a speed same as the background solar
wind (i.e., ambient speed of $\sim$400 {\kmps}) is expected to show a net zero 
acceleration and such an event would take $\sim$103 hours to travel to 1 AU. 
But, the least-square fit curve crosses the zero acceleration line at $\sim$90 
hours and it suggests that on the average CMEs possessed excess energy, which 
supported in expansion as well as propagation. The above equation also 
indicates that events taking travel times $>$90 hours may be accelerated by
the solar wind.  These results are in agreement with Manoharan, (2006).

In Figure 5b, the set of 45 events, which showed estimated initial speed 
greater than the observed speed at the LASCO field of view, are shown by 
over-plotted square symbols. The continuous curve is the
best fit to events shown by square symbols (i.e., {\vest} $>$ {\vint})
and it is given by,
\begin{eqnarray}
 {\rm A_{CME} } = -6.3 + 1.4\times 10^{-1}\, {\rm T_{CME} } - 7.0\times 10^{-4} 
\, {\rm T^{2}_{CME} },
\end{eqnarray}
where A$_{\rm CME}$ and T$_{\rm CME}$ are as given in equation (2). The above 
equation predicts a lesser effective deceleration than the overall curve shown 
in equation (2). Most of these events (refer to events marked by square symbols 
in Figure 5b) show positive and/or null acceleration 
and the decelerating cases on the whole tend to lie close to the zero 
deceleration line (i.e., within about -3 {\mpsq}). For these events, the 
observed travel time and the speed of the CME at 1 AU have revealed the possible 
associated excess of internal magnetic energy (refer to Section 4) in supporting 
the propagation. It is likely that the available energy within the CME aids to 
limit the deceleration in the Sun-Earth distance. It will be interesting to know 
the source characteristics of these events on the Sun as well as their physical
properties at large heliocentric distances (i.e., $>$1 AU). However, points lying 
above the continuous curve have gone through heavy net deceleration in the Sun-Earth 
distance. The investigation of absence or considerably less internal energy of these 
events is also equally important to understand the limited `free magnetic energy' available 
to drive a CME into the interplanetary medium.

\begin{figure}[t]
\centerline{\includegraphics[width=0.5\textwidth,clip=,angle=-90]{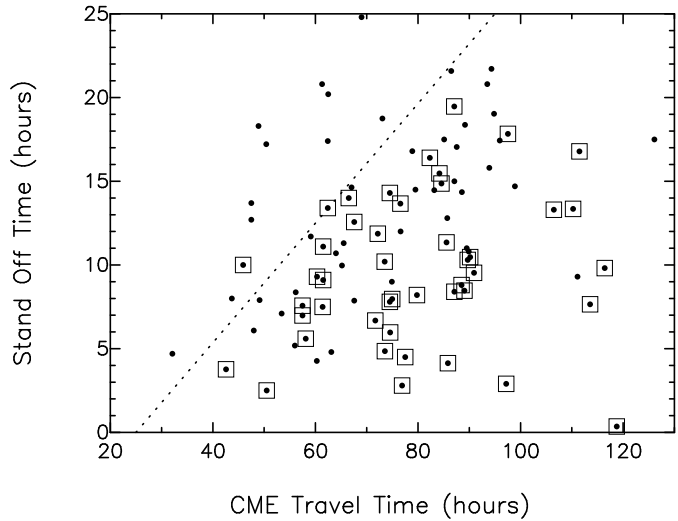}}
\caption{The stand-off time of CME events are compared with their travel
time between the LASCO field of view and Earth. The over-plotted square
symbol shows event for which the estimated initial speed ({\vest}) is larger 
than the observed initial speed ({\vint}). }
\end{figure}

\subsection{Stand-Off Time}

The stand off time of a CME at a distance is given by the time difference 
between the arrival of CME-driven shock disturbance and the CME at the 
given distance. Figure 6 shows the stand-off time plotted as a function of 
travel time of CME to 1 AU. For a given transit time, we see a large scatter 
between 1 to about 25 hours. In general, a smaller stand-off time indicates 
that the driver CME and its shock travel closely together. Whereas a large 
stand-off time may result as the consequence of CME and shock getting apart 
due to the less amount of energy available to drive and/or sustain the shock 
in front of the CME. In the above figure, even at about 40-hours of travel 
time, the stand-off time ranges between $\sim$5 and 20 hours. 

In Figure 6, the events having possible excess internal energy (i.e., events
of {\vest} $>$ {\vint})
are shown by
over-plotted square symbols. The envelop of these events (indicated by a 
dotted line) at the quickly 
transiting side shows small stand-off times of $\leq$10 hours at the Earth (i.e., 
T$_{\rm CME}$ $<$ 50 hours). Those events travel longer to reach 1 AU (i.e.,
T$_{\rm CME}$ $>$ 70 hours) show stand-off times $\leq$20 hours. However,
the scatter in the stand-off time with respect to the travel time (and the
CME speed at 1 AU, {\vicme}) is large, which requires more investigations.

\section{Summary}

In this study, we have studied 91 earth-directed CME events analyzed by 
Manoharan et al.,~(2004) and Gopalswamy and Xie (2008). These two lists 
of halo and partial halo CMEs provide a good sample of events, covering a 
wide range of speeds ($\sim$100--2500 {\kmps}) in the LASCO field of view. 
Each event's interplanetary shock and CME have been detected by near-Earth 
spacecraft. The comparison of the initial speed of the CME with the shock 
and CME speeds at 1 AU shows that every event experiences the aero-dynamical 
drag force in the background solar wind and such drag tends to equalize the
speed of the CME to that of the solar wind. However, some of the interacting 
CMEs take longer travel times than that inferred from their final speeds at
1 AU. The interaction tends to slow down the CME.

For about $\sim$50\% of the events (i.e., {\vest}~$>$~{\vint}; 45 out of 91 
events), the travel time 
to 1 AU, combined with the CME's final speed at 1 AU, reveal an effective 
acceleration in the Sun-Earth distance. This result shows evidence that 
there is a net supply of energy by the CME to accelerate and support the
CME propagation. When the speed associated with a CME is larger than the 
ambient solar wind, the energy of the CME dominates the influences of the
acceleration and/or drag by the background solar wind. The present study
suggests an average energy dissipated per event amounts to be about 
$\sim$10$^{31-32}$ ergs. It is likely that the Lorentz force acting on the 
CME helps to overcome the drag and to support propagation. However, the amount of 
energy transferred from the CME to solar wind over a distance range (or 
a time period) may differ from one event to the other and it is likely 
determined by the physical properties of the CME and
the local solar wind conditions encountered by the CME on 
its way. These results are consistent with the two-step deceleration of
CME events, i.e., a low or moderate deceleration within $\sim$0.5 AU and
a rapid deceleration beginning at around $\sim$0.5 AU (Manoharan et 
al.~2001; Manoharan, 2006). Manoharan (2006) demonstrated that the
internal energy of the CME contributes to the propagation up to about
0.5 AU. 

However, those events experienced net deceleration in the Sun-Earth 
distance (i.e, for events {\vest} $<$ {\vint}), the driving force is 
likely absent and/or not significant. Therefore, it is essential to 
investigate the physical differences between events dominated by the
Lorentz force and those ones without it, respectively, on the Sun and
in the interplanetary medium.

\section*{Acknowledgements}
This work is partially supported by CAWSES-India Phase II program,
which is sponsored by the Indian Space Research Organisation (ISRO).
A.M.R.~acknowledges the facilities and support provided at the Radio 
Astronomy Centre and he is supported by the FIP Programme 
(no. F.ETFTNMK040) of the University Grants Commission, New Delhi. 
We thank G.~Agalya for proof reading the manuscript. SOHO (LASCO data) 
is a project of international collaboration between 
ESA and NASA. 




\end{document}